\documentclass[conference]{IEEEtran}
\IEEEoverridecommandlockouts

\usepackage{cite}
\usepackage{amsmath,amssymb,amsfonts}
\usepackage{algorithmic}
\usepackage{graphicx}
\usepackage{textcomp}
\usepackage{xcolor}
\def\BibTeX{{\rm B\kern-.05em{\sc i\kern-.025em b}\kern-.08em
    T\kern-.1667em\lower.7ex\hbox{E}\kern-.125emX}}
\begin{document}

\title{ML-Enabled Outdoor User Positioning in 5G NR Systems via Uplink SRS Channel Estimates\\

\thanks{This work is partially sponsored by the Swedish Foundation for Strategic Research and Ericsson AB.}
}

\author{\IEEEauthorblockN{Andre Ráth, Dino Pjanić}
\IEEEauthorblockA{\textit{Dept. of Electrical and Information Techn.} \\
\textit{Lund University}\\
Lund, Sweden \\
andrerath23@gmail.com\\
dino.pjanic@ericsson.com}\\
\and
\IEEEauthorblockN{Bo Bernhardsson}
\IEEEauthorblockA{\textit{Dept. of Automatic Control}\\
\textit{Lund University}\\
Lund, Sweden \\
bo.bernhardsson@control.lth.se}\\
\and
\IEEEauthorblockN{Fredrik Tufvesson}
\IEEEauthorblockA{\textit{Dept. of Electrical and Information Techn.}\\
\textit{Lund University}\\
Lund, Sweden \\
fredrik.tufvesson@eit.lth.se}\\
}

\maketitle

\begin{abstract}
Cellular user positioning is a promising service provided by Fifth Generation New Radio (5G NR) networks. Besides, Machine Learning (ML) techniques are foreseen to become an integrated part of 5G NR systems improving radio performance and reducing complexity. In this paper, we investigate ML techniques for positioning using 5G NR fingerprints consisting of uplink channel estimates from the physical layer channel. We show that it is possible to use Sounding Reference Signals (SRS) channel fingerprints to provide sufficient data to infer user position. Furthermore, we show that small fully-connected moderately Deep Neural Networks, even when applied to very sparse SRS data, can achieve successful outdoor user positioning with meter-level accuracy in a commercial 5G environment.
\end{abstract}


\section{Introduction}
\label{section:Intro}
For many years, User Equipment (UE) positioning has been accomplished with Global Navigation Satellite Systems (GNSS), assisted by cellular networks. Besides aiming to achieve reliable and low-latency wireless connectivity, high-accuracy positioning enabled through 5G could coexist and complement existing GNSS-based systems on 5G-capable smart devices. However, GNSS technology is based on unicast transmission and user position is not directly accessible by cellular networks. The latest features within 5G beam forming technologies drive a distinctive need to acquire accurate user location via radio access interface for location-dependent network functionalities such as beam forming algorithms etc. It is expected that in dense urban area deployments, sub-meter mean positioning accuracy can be achieved \cite{Fredrik}, \cite{5GPositioningDLvsUL}. New 3GPP releases are expected to further specify methods for sub-meter accuracy \cite{5GPositioningOverview}. A range of positioning methods, both downlink (DL)-based and uplink (UL)-based, are used. For radio-based positioning,  there is typically a need for specific signals on which a receiver can measure/estimate channel characteristics of interest. This is often expressed as \emph{channel sounding}. Channel State Information (CSI) for the operation of massive multi-antenna schemes can be obtained by the feedback of CSI reports. In a TDD system, the UL channel can be estimated based on SRS transmitted from each UE for which the base station (BS) estimates the DL channel by exploiting channel reciprocity \cite{Marzetta}, \cite{38211}. UL channel estimation includes estimating the Time of Arrival (ToA), the received power, and the Angle of Arrival (AoA) - all being parameters from which the position of the User Equipment (UE) can be estimated. As defined in 3GPP \cite{38211}, the SRS is a UL Orthogonal Frequency Division Multiplexing (OFDM) symbol with a Zadoff–Chu sequence on its subcarriers, known by both the UE and BS.
\newline
\indent Positioning by radio signals is enabled through methods such as fingerprinting or model-based estimation using signal features \cite{VTC}. The multi-path information of the environment is embedded in the CSI data, and hence the CSI can be used to characterize the radio environment. Examples of CSI-based indoor positioning were presented in \cite{CSI},\cite{CSI2} while researchers in \cite{CoDesign_MMWave} and \cite{RayTracingML} demonstrated UE positioning via beam information from Reference Signal Received Power (RSRP). The work presented in \cite{RussLTE} shows that the statistics of the wireless channel in Long Term Evolution (LTE) can be used to create a positioning solution even in non Line-of-Sight (NLoS) conditions through an azimuthal-delay representation of the wireless channel. Another LTE DL reference signal-based approach \cite{RussNLOSNavi} demonstrates that multipath effects can be utilized advantageously to estimate not only user position but also orientation through wireless fingerprinting. 

In related literature, spatial fingerprinting in conjunction with classical machine learning (ML) methods enables UE localization via learned features of the environment \cite{PhasedPos}. Recent positioning-related results in \cite{UDNPosition}, \cite{AccurateVehcLocalization}, also applicable to mmWave networks, target localization accuracy in cases where either the network is optimized for positioning applications or the positioning algorithm is tailored to the particular network geometry. One of the few studies exploring a UL-based method is \cite{SRS} where simulated UL SRS channel estimates are utilized to investigate the feasibility of SRS estimates for 3D positioning based on joint angle-time estimation and expectation-maximization. Another UL-based method was presented in \cite{SRS_LTE}, where indoor positioning through simulated UL SRS signals in LTE-FDD was presented. While the vast majority of the studies above rely on various DL-based methods for user positioning, we opt to demonstrate a novel ML-powered method using UL channel estimates from SRS transmissions generated in a real-world 5G base station (gNodeB). To the best of the authors' knowledge, there is no prior work on this matter. The main contributions of this paper can be summarized as: \newline
\indent We demonstrate that UL SRS-obtained channel estimation in the BS provides sufficient information to regress for UE position through Deep Neural Networks (DNN). In this study, we consider sparsity to be defined as using channel estimate information from only a small fraction of the available Physical Resource Blocks (PRBs). Data sparsity enables minimal data processing overhead and the use of DNNs that are both low-power and moderately shallow, which reduce the risk of causing potential delays and capacity overloads, necessary for real-time Layer 1 processing where decisions must be made in milliseconds. \newline 
\indent Furthermore, in contrast to the majority of prior studies, we prove the viability of SRSs collected in a real commercial 5G NR network setup instead of a simulated environment or non-commercial setup. From a technology perspective, ML is about improving network decision-making capability and allowing it to learn from patterns \cite{EricssonHowAndWhyRANwithAIML_5G}. 
The latter urges designing ML-powered methods for real-time operations with the capability to solve complex and unstructured problems using data collected at L1-L2 interaction between BS and UE. Decisions need to be made near where data is generated \cite{EricssonRelationAI_5G}. 
\section{Data Collection And Methodology}
\begin{figure}[!b]
    \vspace*{-2mm}
    \centering
    \includegraphics[width=0.88\linewidth]{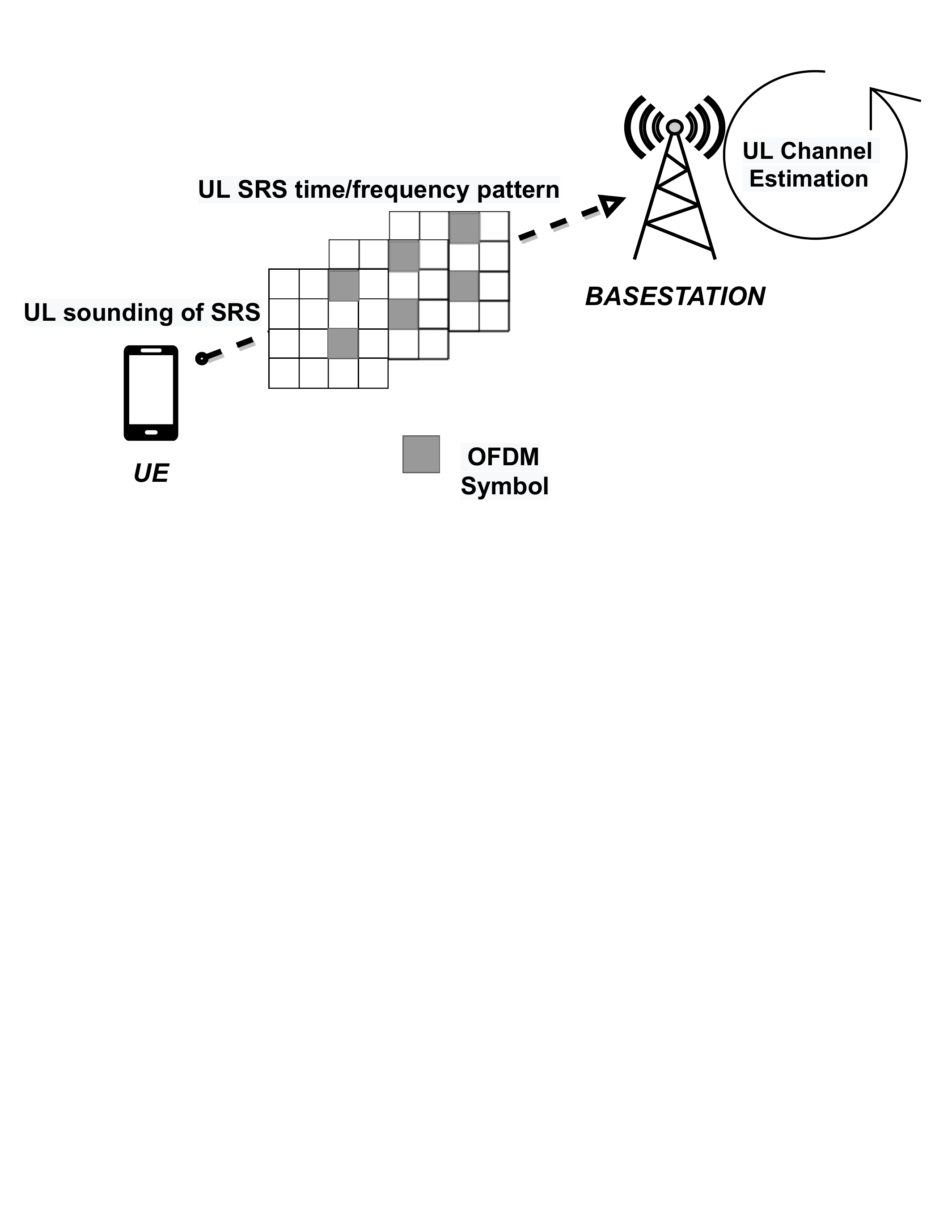}
    \vspace*{-60mm}
    \caption{UL SRS transmission from a UE; The BS obtains Sounding Reference Signals (SRS) containing channel information data from the UE. The SRS is designed to cover the full bandwidth, where the resource elements are spread across the different symbols to cover all sub-carriers. Therefore, SRS is designed with a comb-based pattern.}
    \label{fig:SRS_feedback}
    \vspace*{-2mm}
\end{figure}

\noindent To establish an ML-driven proof-of-concept (POC) for positioning using channel estimates from UL physical layer (L1) channel SRSs, we employ a commercial-grade 5G radio system compliant with 5G NR 3GPP 38.104 Rel15 \cite{38104}. A commercial-grade Phased Array Antenna Module (PAAM) is utilized in a BS operating at the center frequency of 3.85~GHz with a 100 MHz bandwidth. We used a proprietary 5G-capable, Android-based test UE, with user motion in different mobility patterns at a distance of approximately 70 m from the roof-top antenna.  We opt to extract the channel estimates from the BS baseband unit, which processes the time-varying SRS reports as per the SRS feedback loop structure depicted in Fig. 
 \ref{fig:SRS_feedback}. 

The general thought behind positioning with UL channel estimates is that a physical location under similar network conditions roughly corresponds to a specific SRS-generated channel matrix estimate. In other words, different locations in space have distinct channel fingerprints. 

Continuous data collection was specifically chosen for this study to mirror realistic navigation conditions. The SRSs are designed to cover the full bandwidth, where the resource elements are spread across the different symbols to cover all subcarriers. In the proprietary baseband hardware unit, the internal beam-space representation of the channel can be extracted and post-processed after ensuring that the UE had high data-rate signalling throughout the channel measurements through 4K video streaming.
The SRS-derived channel estimates are stored in a complex-valued matrix structure, which henceforth is referred to as a \textit{channel matrix}. For every SRS sampled from a specific UE, the BS channel matrix contains channel estimates of 64 directional BS antenna elements (directions) for each UE antenna and PRB container. The test UE supports a 1/2/4-antenna configuration, of which the 4-antenna configuration was used during our testing. Furthermore, the 100 MHz Time-Division Multiplexing (TDD) configuration supports 273 PRBs, which in the BS are then allocated to containers with a configurable number of 2, 4, or 8 PRBs per container. In our setup, 2 PRBs were enabled per container. Therefore, there were 137 frequency channels configured, each containing two adjacent PRBs. The channel matrices retrievable from the BS thus contained one complex value/antenna direction for every PRB container and UE/BS antenna pair. In summary, the retrievable channel estimate $\mathbf{H}$ from our experimental setup consists of a complex-valued matrix with up to 137 frequency channels, 64 BS antennas and 4 UE antennas. Together, the upper limit for data extraction in our experimental setup consists of 35072 complex values per SRS transmission. 
\begin{equation}
 Max\left[\text{Cap}\left(\mathbf{H}\right)\right] =  Max\left[N_{Ch}N_{tx}N_{Dir}\right] = 137\cdot 4 \cdot 64,
   \label{eq:MEAS_dataCount}
\end{equation}
where $N_{Ch}$ is the number of channels, $N_{tx}$ is the number of UE antennas and $N_{Dir}$ number of BS antenna elements. With 35072 complex values extracted potentially every few milliseconds, the internal data amount handled becomes a major concern. Due to data rate constraints during data logging, we first aimed to explore how a small data amount is sufficient for meter-level positioning. Only 3 containers with 2 PRBs each were retrieved, hence 768 input features, or in other words, 12 of what we term \textit{sub-channel matrices}, denoted as $\textbf{H}_{8\times8}$. Using only a sparse 768 of the potential 35072 values does not prevent positional information from being extractable from the SRS-obtained channel estimate. As shown later in the paper, 768 features still present a unique opportunity for ML frameworks to learn and later regress for the UE position. Furthermore, since SRS measurements are periodic for a given 5G NR waveform numerology, they present an ideal opportunity for ML frameworks to utilize for UE localization. The resulting single-measurement data matrix had a dimensionality as follows: 
    \begin{equation}
\mathbf{H}_{N_{Ch} \times  N_{TX} \times N_{\psi,h}\times N_{\psi,v}} =  \mathbf{H}_{3,4,8,8}.      \label{eq:Dimensionality}
    \end{equation} 
Channel matrices may also be expressed as in
(\ref{eq:Dimensionality_fold}), 
    \begin{equation}
\mathbf{H}_{\left(N_{Ch} \cdot  N_{TX} \right)\times N_{\psi,h}\times N_{\psi,v}} =  \mathbf{H}_{12,8,8}.      \label{eq:Dimensionality_fold}
    \end{equation} 
To utilize the entire 100 MHz bandwidth, the three PRB containers chosen were the lowest, middle, and highest sub-channels. This represents solely 2 $\%$ of the total number of possible sub-carriers. The final logging aspect is that of time. In this case, two timestamps are logged: the frame number corresponding to the actual network time, and the UTC time-stamp corresponding to the time a given SRS dataset was collected and written to the log, used to regress UTC-timestamped GNSS position to channel fingerprints.  

\begin{figure}[ht]
\vspace*{-4mm}
    \centering
    \centerline{\includegraphics[width=\columnwidth]{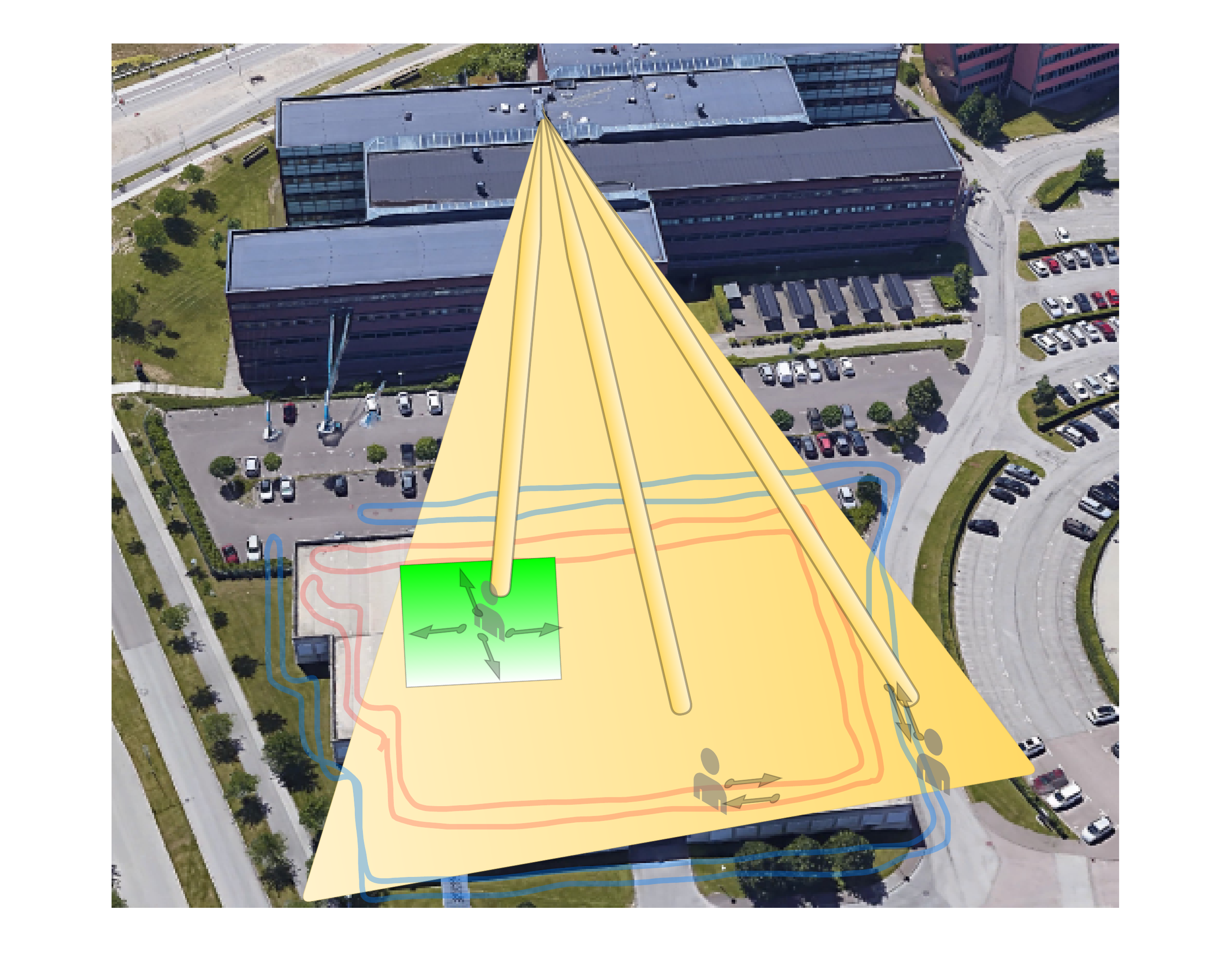}} \vspace*{-3.5mm}
    \caption{The pre-defined measurement routes in a SU-MIMO scenario: A 2-story, 10 m high garage building where the red line on the top of the building is the LoS route. The blue line is representing the ground-level route where the surrounding buildings block and reflect the signal from a 20 m high rooftop antenna causing NLoS propagation. The vivid green square shows the region for the LoS dense-walk route. The general coverage area is illustrated in light yellow color whereas yellow-colored narrow beams were generated by the BS equipped with a 64-antenna element array.}
    \label{fig:MEAS_route}
    \vspace*{-3.5mm}
\end{figure}

\indent To decide the geographic area to conduct the measurement campaign in, a few aspects were considered. First and foremost, as the training process is based on GNSS data in our setup, the GPS signal had to be preferably unobstructed throughout the route. To investigate the validity of the proposed ML approach, both LoS and NLoS scenarios were investigated. Three predefined routes were used as the baseline for positioning: a square-shaped area of the dense walk for the training set with a natural random-walk validation and test dataset, an LoS path training set for positioning in the larger area along a predictable path, and an NLoS path data nearby to compare LoS and NLoS effects, as depicted in Fig. 
 \ref{fig:MEAS_route}. Both the rooftop LoS scenario and the ground-level NLoS scenario, respectively henceforth referred to as LoS-A and NLoS-A. The square-shaped area of the dense walk data will be termed LoS-D. The UE moved at the standardized pedestrian velocity of 3 km/h in all the scenarios. We remark here that our target was to study pedestrian velocities, velocities higher than pedestrian ones were not within the direct scope of this study hence conclusions made here shall not be extrapolated to those.

We collected three distinct datasets for each scenario: training, validation, and test. These datasets were collected in different acquisition sessions with identical measurement setups. In the LoS-D scenario, the training, validation and test dataset collection happened on the same day; for the LoS-A scenario the training dataset collection was done a week before the test and validation dataset collection, which happened on the same day. For the NLoS-A datasets, the training and validation datasets were collected on separate days of the same week, while the test dataset was collected a month later. We would like to emphasize that the test datasets were not touched by either the model or the data processing pipeline before results were evaluated, and were \emph{not} factored in during model selection either. However, manual data analysis and processing were conducted by the authors on the test data \emph{before} model evaluation to examine data validity. 
\section{Data Processing Pipeline}
\subsection{SRS Data}
\begin{figure}[t!]
    \centering
    \includegraphics[width=0.95\linewidth]{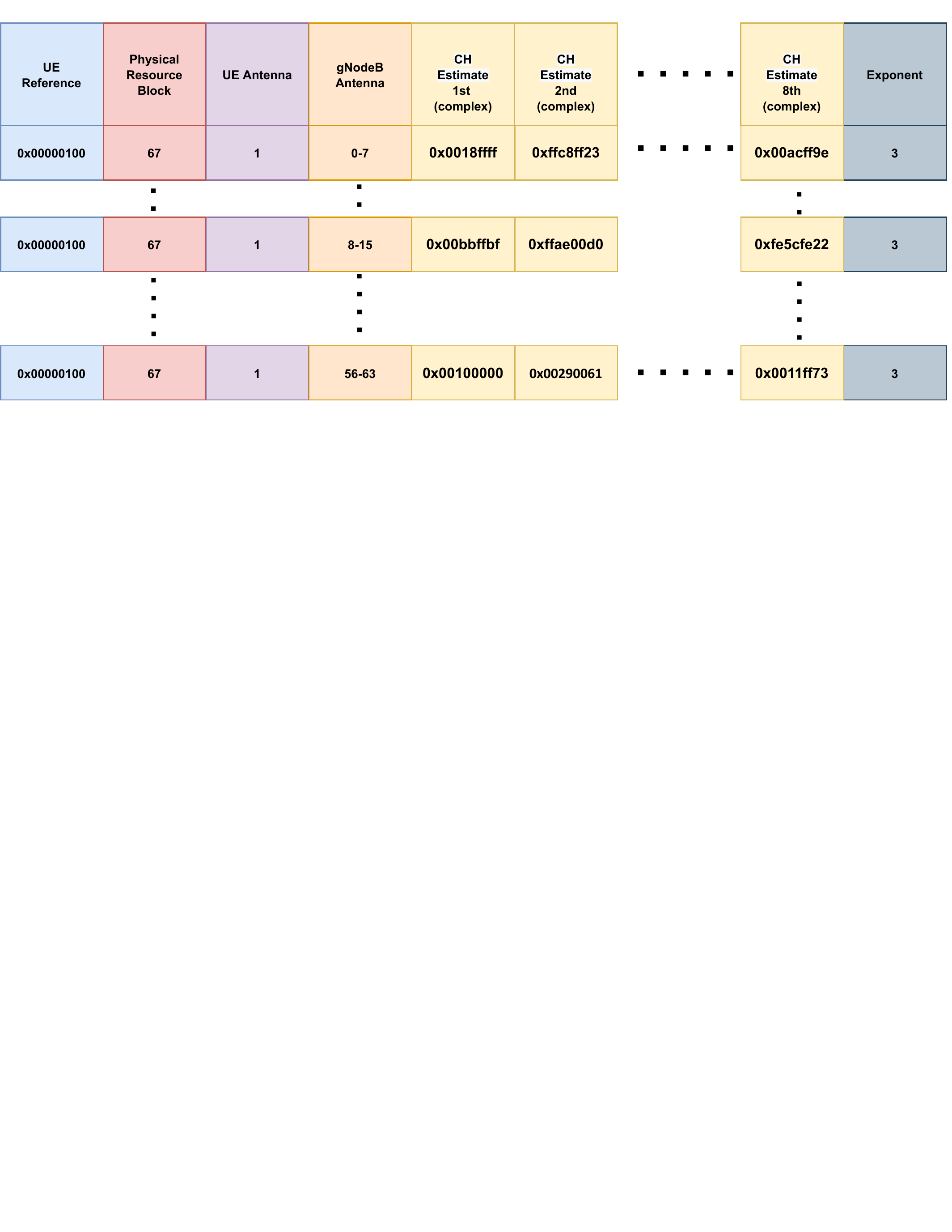}
    \vspace*{-74mm}
    \caption{The format of a single dataset instance, containing the SRS channel estimate for a single frequency channel and UE antenna.}
    \label{fig:MEAS_SRS}
    \vspace*{-3.5mm}
\end{figure}
With the SRS-derived raw dataset obtained from the baseband module in BS, the next step was retrieving the $\mathbf {H}_{12\times8\times8}$ matrices from the data logs. The SRS-derived channel estimate dataset is generated per SRS measurement occasion, meaning per cell, symbol, and UE antenna. It is stored in 1-12 subchannel/UE antenna pair order for all BS antenna directions, down to millisecond intervals. Channel estimates are represented as 4 hex digits for the real and the imaginary components.
A representation of the dataset format can be seen in Fig. \ref{fig:MEAS_SRS}.
\noindent We consider the System Frame Numbers (SFN) numbers from a repeating sequence of 0 to 1023 throughout the measurement. The series of channel matrices $\mathbf {H}_{12\times 8 \times 8}[SFN]$ are then stored in a 4D matrix $\left[\textbf{H}_{12\times8\times8}\right]_{N_{data}}$, where $N_{data}$ is the number of unique SFN during which at least one sub-channel matrix was measured. The first step in feature selection is then separating the phase and amplitude of the complex channel matrices. 
\subsubsection{Phase component} 
Comparing the phase in the extracted data to the phase in the raw channel matrix in our experimental setup, we find that the gNodeB's built-in beam-domain transformation uses the received signal phase. The ML system then obtains data with the angle of departure from the gNodeB already utilized. The remaining information in the phase of the complex numbers in the ML input data is discarded in this work due to its dynamic nature.
\subsubsection{Amplitude component}
\begin{figure}[t]
    \centerline{\includegraphics[width=0.9\linewidth]{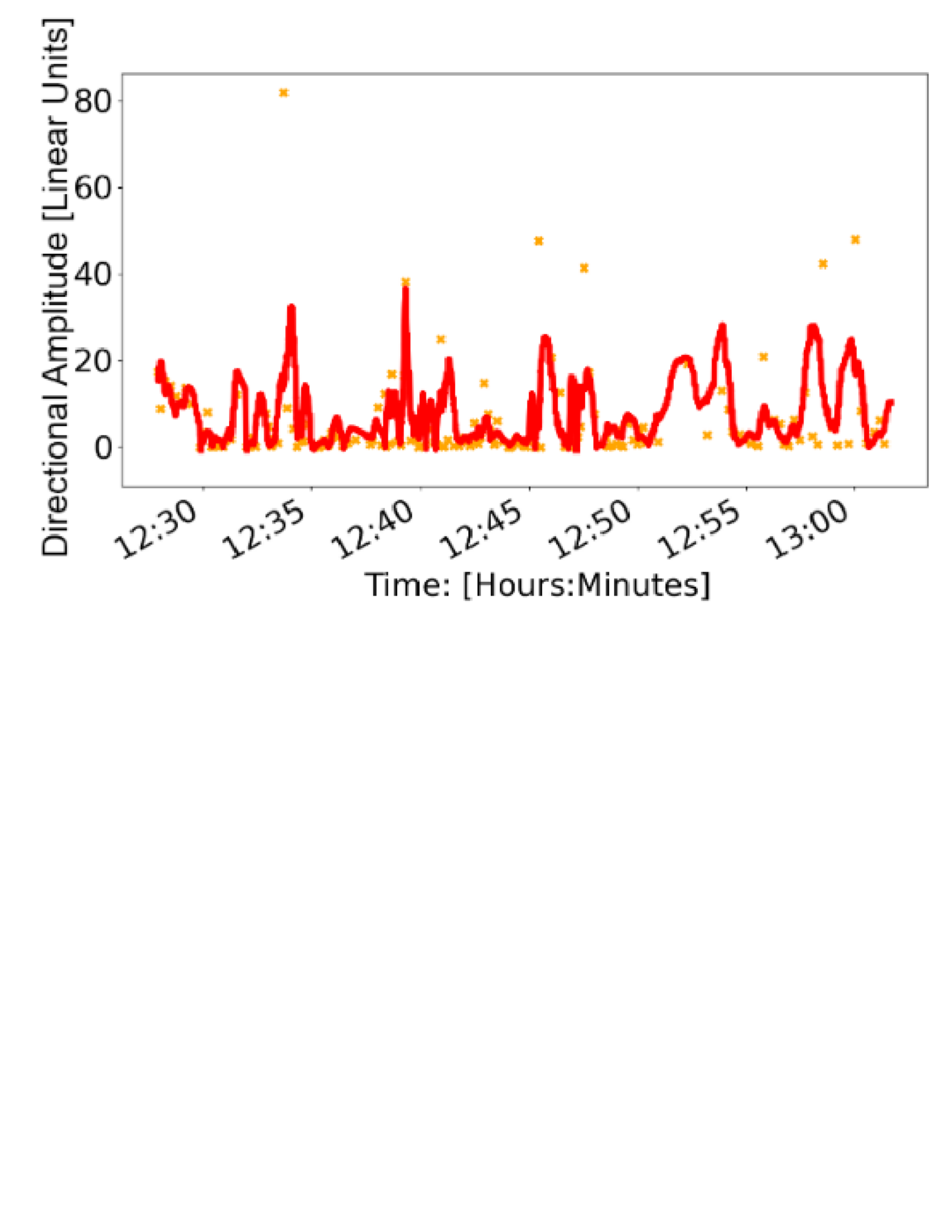}}
        \vspace*{-52mm}
    \caption{A snapshot of the complex amplitude output of a single direction $h_{[8,5]}$ in a sub-channel matrix $\textbf{H}_{i}$ of the NLoS-A1 database. The red curve shows the amplitude smoothed over 100 samples.}
\label{fig:Data_amplitude_period_correlate_unibeam}
\end{figure}

\begin{figure}[t]
    \vspace*{-1.9mm}
    \centerline{\includegraphics[scale=0.26]{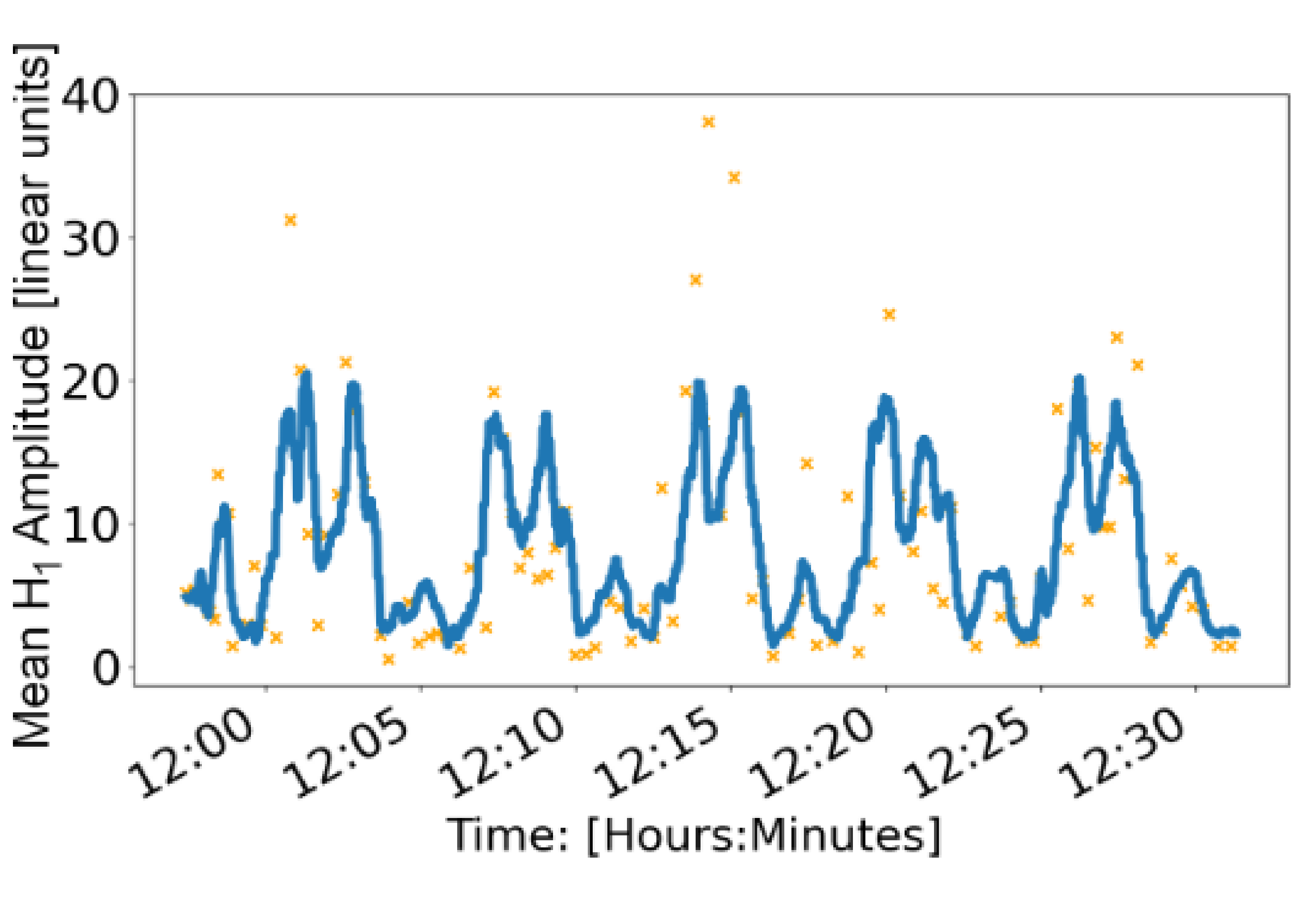}}
      \vspace*{-4mm}
    \caption{The mean complex amplitude of all the outputs in a sub-channel matrix $\textbf{H}_{i}$ of the NLoS-A2 database. The amplitude of the outputs varies over 5 periods with the periodicity expected from the path dataset. The blue curve shows the moving average of the amplitude over 100 samples.}
    \label{fig:Data_amplitude_period_correlate_mean_beam}
    \vspace*{-4mm}
\end{figure}

The extracted data amplitude should contain meter-scale positional information arising from large-scale fading. An underlying assumption is that in both LoS and NLoS cases, the amplitude transfer function of a radio signal depends on environmental geometry, with amplitude thereby acting as a slowly-varying correlate to a position. This will then be visible in the extracted data from the BS - the periodicity of position as the path is walked back-and-forth on is expected to result in a similar periodicity in the recorded complex amplitude. On the NLoS-A dataset, for example, there are 5 back-and-forth cycles on the NLoS path seen in Fig.\ref{fig:MEAS_route}. The expectation is then that certain outputs will have very visible periodicity, e.g. as confirmed in Fig.\ref{fig:Data_amplitude_period_correlate_unibeam}. This becomes more clear when the amplitude for all the outputs in a single sub-channel matrix is examined, as per Fig.\ref{fig:Data_amplitude_period_correlate_mean_beam}.
\subsection{GNSS Data}
A commercial UE was used to record GNSS data with an open-source android app to interface with the Android GNSS API. The app obtained GNSS-INS (Inertial Navigation System \textit{Position Navigation Timing} estimates at a 1 Hz sample rate. The UE model was running OxygenOS 11 with Dual-band Multi-Constellation GNSS rated at 3.5 ± 0.5 (m) horizontal accuracy.

\subsection{Combined Data}
To use the extracted complex-amplitudes of the sparse channel matrices $\left[\textbf{H}_{12\times8\times8}\right]_{N_{data}}$ as input for position regression with ML, further processing is required. As not all sub-channel matrices are updated during SRS transmission, the missing channel estimate values for any given sample time must be somehow represented for the DNN.  

\begin{figure}[!t]
\vspace*{-2mm}
    \centering
    \includegraphics[width=\linewidth]{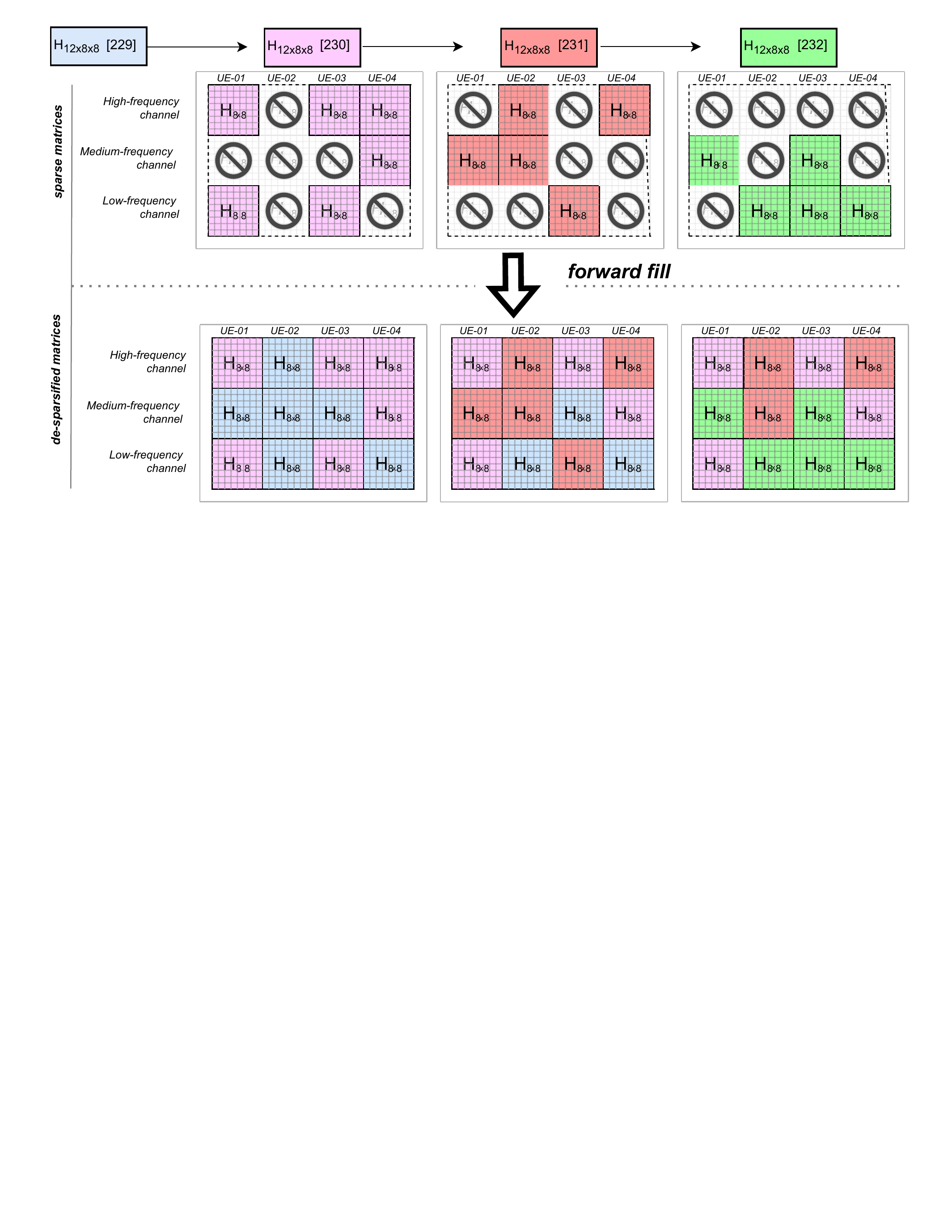}
    \vspace*{-71mm}
    \caption{Using forward-filling on  channel matrices $\mathbf {H}_{12\times 8 \times 8}[SFN]$. For any sub-channel matrix $\mathbf {H}$ at SRS sample-time $\tau$, if a value is not given by the current SRS then the most recent known value for that sub-channel matrix is used instead.}
    \label{fig:Data_ffil}
\end{figure}

Filling in the missing channel-matrix values for any given $\mathbf {H}[\tau]$ extracted channel matrix at sample-time '$\tau$' is the first step in our data pre-processing pipeline. The simplest method for filling in missing data without using future values or known priors is forward-filling the latest known values. Forward-filling for the channel matrix $\mathbf {H}$ is visualized in Fig.  \ref{fig:Data_ffil}. During our measurements, on average 6 of the 12 channel sub-matrices were refreshed every SRS sample, and we observed a variable delay between samples of approx. 35-110 ms, with a higher sample rate in NLoS conditions and a lower sample rate in ideal LoS conditions. This refresh rate was high-enough that unless a connection drop is observed, most sub-channel matrices only persisted for under half a second.

For data normalization in this study, only linear scaling was utilized, with improvements in this area left for future work. This was done using min-0 max-1 scaling of the datasets by simple division. The normalization factor was determined by obtaining the maximum amplitude present in the training data, thereby preventing the contamination of the validation and test datasets. Furthermore, we found that taking the square root and fourth root of the channel matrices substantially improved validation positioning results. The exact cause of this performance improvement is unclear and may be the topic of future investigation. However, the maximal benefit was achieved for NLoS scenarios when both square- and fourth root of the input data was used. For the LoS scenarios, the fourth root was of unclear benefit. 

The square-root and fourth-root concatenation came at the cost of doubling the number of input parameters to the network, to 1524 in total. However, this many parameters as input for a fully-connected network could lead to overfitting. From the assumption that location is mostly independent of UE orientation, reducing the number of input parameters can be achieved with low performance penalty by only taking one $\mathbf{H}_{i}$ sub-channel-matrix-equivalent as input for every sampled frequency channel. With an eye for future scalability w.r.t. different configured UE antenna numbers, this would also enable ML systems input parameter counts to be independent of different UE antenna configurations and models. 

For this reason, we take the average per direction of the sub-channel-matrices $\mathbf{H}_{i}$ belonging to the same frequency channel. This enables dimensionality reduction without losing frequency-channel information: (\ref{eq:per_diection_channel_mean}). \vspace*{-0.5mm}
\begin{equation} \vspace*{-0.5mm}
 \mathbf{H}^{mUE}_{3\times 8 \times 8} = \frac{1}{N_{TX}}\sum^{N_{TX}}_{i=1} \mathbf{H}^{\text{UE}_{\text{antenna}}=i}_{N_{Ch}\times N_{\psi,h}\times N_{\psi,v}}
   \label{eq:per_diection_channel_mean}
\end{equation}

\begin{figure}[t!]
\vspace*{-2mm}
\centering
\includegraphics[width=0.92\linewidth]{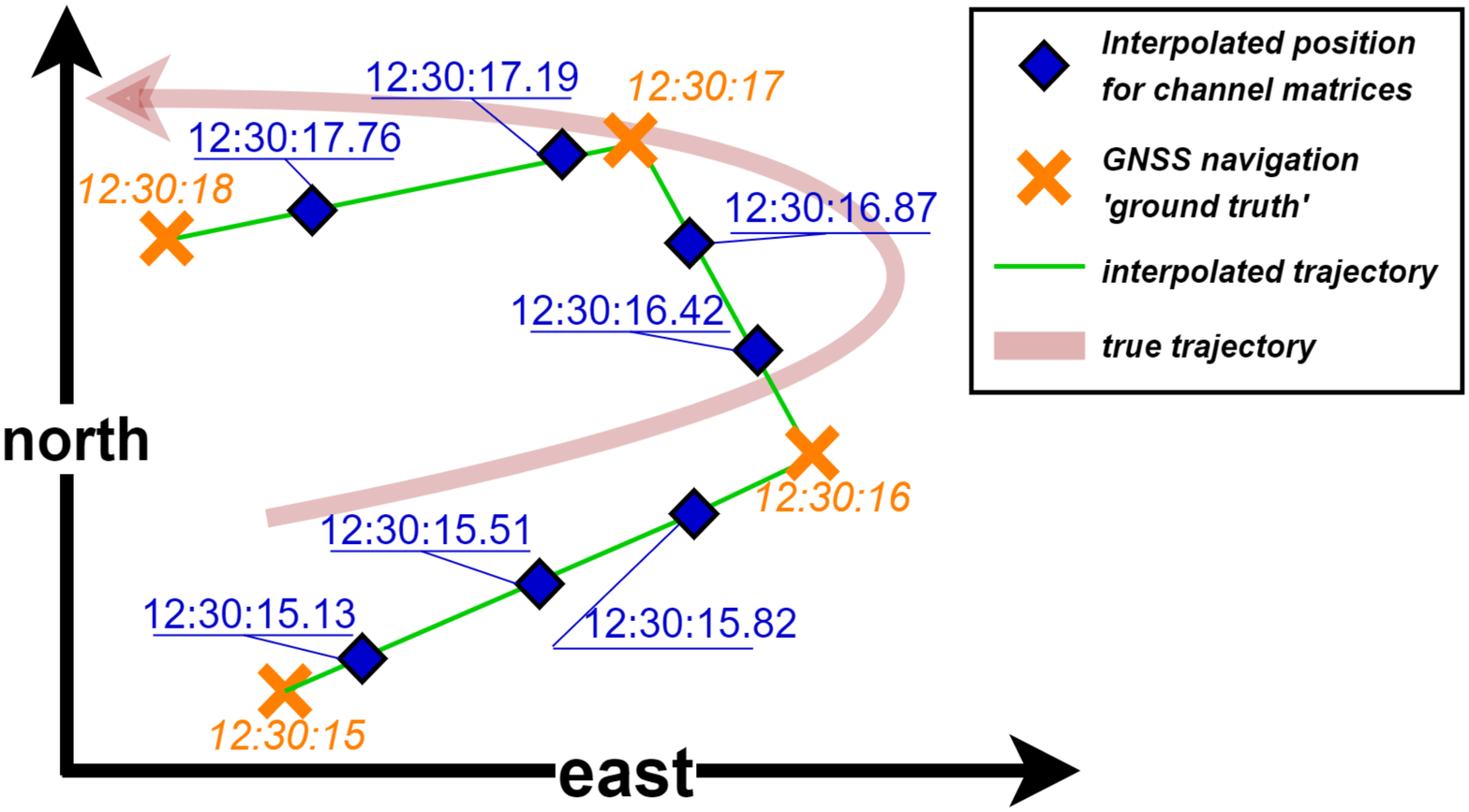}
\vspace*{-14mm}
\caption{Assigning position to channel matrices $\mathbf {H}_{12\times 8 \times 8}[SFN]$ using shared UTC timestamps with the GNSS dataset and simple linear interpolation. Also shown is the GNSS-based navigation position fix deviation from the 'true' pedestrian trajectory.}
\vspace*{-4mm}

\label{fig:Data_GNSS_interpolation}
\end{figure}
To assign positioning \emph{ground truths} to the channel matrix data $ \mathbf{H}^{mUE}_{3\times 8 \times 8}$, the UTC timestamp of both the SRS Channel matrix data and the UE position output is used. First, the two datasets are synchronized. Linear time-interpolation from the GNSS-position data is used to create interpolated trajectories, through which the 'ground truth' $P_{EN}$ coordinate pairs for each channel matrix is generated, which can be converted to local $P_{XY}$ coordinates. Finally, all channel matrices that fall outside the bounds of the GPS measurement are discarded. The position interpolation process is shown in Fig.  \ref{fig:Data_GNSS_interpolation}. GNSS inaccuracy is partially modelled during the training process by injecting Gaussian noise of similar magnitude as the GNSS measurement onto the training data $
P_{train}$ every epoch during the DNN training process. This also functions as output regularization.
\section{Proposed Machine Learning Framework} \label{MLModels}

\begin{figure}[t]
    \centering
    \includegraphics[width=0.95\linewidth]{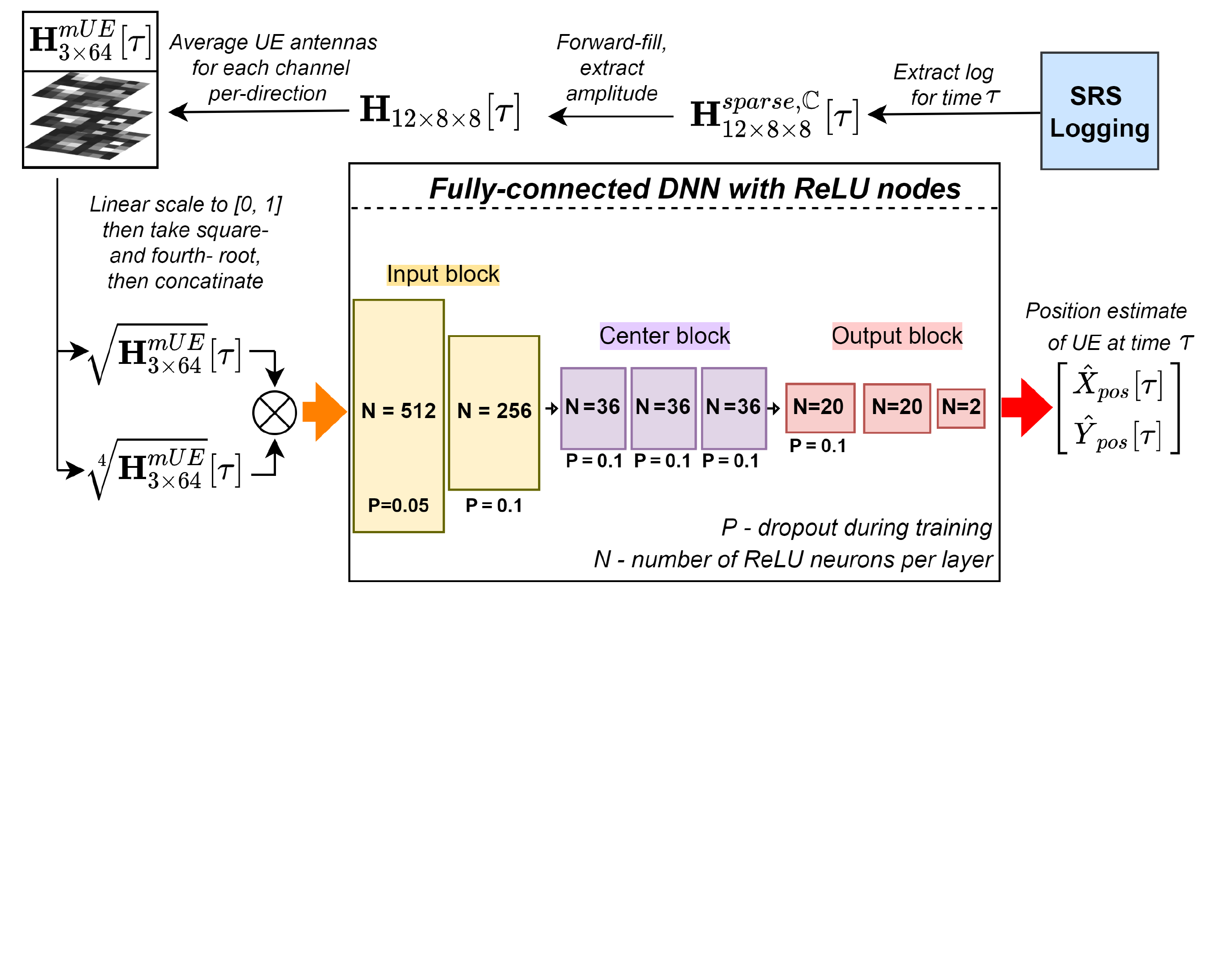}
    \vspace*{-24mm}
    \caption{The 3-block fully connected DNN used in this study along with the input pipeline and intended output. The per-layer dropout used while training is also shown.}
    \label{fig:MLMethods_blocks}
\end{figure}

In this section, we describe the network architecture used for learning and discuss some design aspects. The proposed system architecture is illustrated in Fig.  \ref{fig:MLMethods_blocks}, where the fully-connected Deep Neural Network (DNN) uses features from the SRS dataset as input to regress for local $P_{XY}$ position. 

To demonstrate a real-time POC ML-driven positioning with minimal computing overhead, only small moderately deep fully-connected DNNs were tested. Architectures with up to 15 layers at a maximum of 128 artificial neurons (ANs) per layer after the input and first hidden layers were non-comprehensively searched. Of the tested DNNs, the best-performing architecture on the validation data was selected, with no tuning or selection done using test data. Network architecture was unchanged between the ML models for the different datasets. For the final hyperparameter search, three discrete blocks of fully-connected layers were defined, each with the rectified linear unit (ReLU) activation functions and varying AN and layer counts within a limited range. 
\begin{enumerate}
\item \textbf{Input block:} The \emph{input block} serves to take the input data through subsequent shrinking layers into a parameter bottleneck, compressing the data. 

\item \textbf{Center block:} The expectation is that the process block takes the reduced dimensions from the input block and feeds it through identical fully-connected layers, processing the lower-dimensional representation further. 

\item \textbf{Positioning block:} The expectation is that the positioning block takes the center block's output and finally narrows it down to two dimensions to regress for a position. Note that the last layer has two outputs corresponding to a position's local X and Y coordinate pair ($P_{XY}$).
\end{enumerate}
This design choice originated from our testing where introducing a bottleneck of 20-40 fully-connected ANs per layer for all but the first two layers reduced overfitting while having a wide input and first hidden layer improved general performance. The overall number of hidden layers was also kept low, as increasing layer counts over 7 did not discernibly affect validation performance. 

Finally, the  minimized loss for the network is the Mean Euclidean Distance Loss (MEDL), which can be expressed as: \vspace*{-2mm}
\begin{equation} 
    \text{MEDL} = \frac{1}{N_{s}}\sum^{N_s}_{\tau=1} \lVert P_{XY}[\tau] - f_{\theta}\left( \mathbf{H}^{mUE}\left[\tau\right]\right) \rVert_1,\\
\end{equation}
where $f_{\theta}$ is the ML model with $\theta$ optimizable parameters, $P_{XY}[\tau]$ the interpolated GNSS coordinates for sample-time $\tau$, and $N_s$ is the number of time-samples in a batch. The loss was minimized with an ADAM \cite{ADAM} optimizer using PyTorch on a CUDA-capable GTX 2060.

\section{Results And Discussion}
\vspace*{-0.5mm}

To summarize our results shown in Table \ref{tab2}, we obtain an approximate mean euclidean distance of 3-9 m as compared to the GNSS data when evaluated on test data, with accuracy depending mostly on data conditions. We note again that model selection or parameter optimization was not done on test data. During validation, it became apparent that data character changed between the training datasets in LoS-A and the validation datasets, potentially explaining the degraded performance compared to LoS-D, where no domain change was observed.
\begin{table}[h]
\vspace*{-3mm}

\caption{Mean euclidean error in meters for each dataset}

\centering
\vspace*{-2mm}

\begin{tabular}{c||c|c}

    dataset name & validation dataset & test dataset  \\
            \hline
            LoS-D   & 2.8 (m) & 3.3 (m) \\ 
    LoS-A   &  9.2 (m) & 9.7 (m) \\ 

    NLoS-A  &  7.3 (m) &8.1 (m)\\ 

        \hline
            \end{tabular}
    \label{tab2}
    \vspace*{-1mm}

\end{table}

These results compare favourably to results found in the literature on most outdoor positioning systems using similar DL/UL-based positioning approaches and density real-world data. The NLoS data accuracy indicates this method's viability for positioning in a real-world environment. We note that the precise effects on the SRS channel matrices of non-pedestrian tracking at high velocity and, e.g., users in a vehicle have not been tested. As an example, forward-filling introduces data from previous sample times. For scenarios where significant distance may be travelled between SRS samples, alternatives to forward-filling might be needed e.g. using only the latest SRS as a partial data point. As an extension of this work, we show a proof-of-concept in \cite{AndreRath_AI} where the same datasets and ML pipeline introduced in this paper are extended with simulating pedestrian motion through particle filtering, improving mean accuracy to around 5-6 m for NLoS scenarios. \newline
\indent To summarize, this study demonstrates the practical viability of UL SRS channel estimates in a realistic outdoor NLoS propagation environment. In contrast to other studies employing multi-antenna arrays at the receiver side, we use a commercial-grade, 4-antenna-equipped UE. 
\section{Conclusions And Future Work} \vspace*{-0.2mm}
We have shown some of the potentials of DNNs for outdoor user positioning in 5G NR systems using UL SRS channel estimates in a very sparse data sampling regime. The results presented show sub-10 m of mean accuracy for all test scenarios, despite an already inherent ground-truth horizontal positioning inaccuracy of 3.5 ± 0.5 m in the GNSS dataset. A more accurate GNSS positioning setup for training data should substantially improve results. Similarly, a higher SRS sampling rate should also improve the positioning results significantly. For future research, the phase of the SRS channel estimates could be a possible feature source to explore. Finally, considering the simplicity of the DNN model we used could also be interesting, as more sophisticated models may further improve accuracy.

\end{document}